# Energy-Efficiency Routing algorithms in Wireless Sensor Networks: a Survey

Ghassan Samara, Ghadeer AlBesani, Mohammad Alauthman, Mohammad Al Khaldy

**Abstract**— A Wireless Sensor Network (WSN) is a collection of tiny nodes that have low energy levels and have become an essential component of the modern communication infrastructure and very important in industry and academia. Energy is crucial in WSN, and thus the design of WSN in the research community is based on energy efficiency, and node energy consumption is a great challenge to enhance WSN's lifetime. It may be costly or even impossible to charge or replace consumed batteries because of the difficult environment. Many energy efficiency methods are introduced in this article to decrease energy consumption, improve network performance and increase network lifetime.

———————————————— ◆ ————————————————

## 1. INTRODUCTION

A Wireless Sensor Network is a collection of tiny power, multifunctional and communications nods with observation and recording situations at distinct places, afterwards, convert this data to signals that can be processed, Such nodes are randomly implemented on a large or small scale, this becomes a significant field for study because these networks are used today in numerous consumer and industrial applications, for instance in healthcare, the industry, the transport system, government security and military systems, the environment and agriculture and underwater sensor systems. If the amount of sensors is big, this enables for greater monitoring with greater accuracy, but it can be very costly or even impossible to charge or replace batteries because of the challenging environment. These scattered sensor nodes are capable of collecting and transferring data back to an internal base station (BS) or other sensors, sending and receiving information drain node's energy, Therefore, the best way to improve the life of WSN is by selecting information transfer paths to minimize the complete drainage of energy along the route and to balance the load between the nodes. The BS can be either a mobile or a fixed node that connects the sensor network to a current infrastructure for communication or the Internet. Since the WSNs have become an important element of the modern infrastructure of communication for the 21st century [1] [2], In order to effectively provide information to their destination, the power consumption and maximise network life have become the critical parameter characteristic in routing protocols.

Tackling energy efficiency issues should be done by taking the application specifications into account, so the choice of routing protocols depends on application specifications and network architecture. Applications activate nodes for a long time so that in a few days the node loses its battery. This problem has resulted in protocols that can decrease energy drainage. WSN routing is a very difficult and extra burden, which can lead to severe energy consumption. Energy-constrained sensors are anticipated to operate independently for an extended period of time. WSN designers cannot therefore easily select an effective protocol to save node energy while preserving the desired network operation, the reasons for potential energy waste, sensor drainage during the detection, processing, transmission or reception of information are presented in this paper in order to accomplish the job anticipated from the implementation. In terms of communication, there is also a good deal of energy consumption, like

**Collision:** which happens if a node receives a lot of packets simultaneously. All packets that caused the collision should be ignored and retransmitted.

**Overhearing**: Nodes in the sender's transmission spectrum will receive the packet sent, even if the packet is not for it, so energy is consumed unnecessarily.

**Idle listening:** One significant reason for the waste of energy is when the node listens to an inactive traffic channel..

## 2. LITERATURE REVIEW

### 2.1 Energy Consumption

Many works are underway to optimise the use of energy in battery-limited sensor systems. These protocols are created by taking into account the application demands and network architecture. However, when designing WSN routing protocols some factors should be considered. The key factor is the sensor's energy efficiency, which has an effect

————————————————
- *Ghassan Samara, Department of Computer Science, Faculty of Information Technology, Zarqa University, Zarqa, Jordan Email : gsamara@zu.edu.jo*
- *Ghadeer Albesani Department of Computer Science, Faculty of Information Technology, Zarqa University, Zarqa, Jordan.Email: ghbesani@zu.edu.jo*
- *Mohammad Alauthman, Department of Computer Science, Faculty of Information Technology, Zarqa University, Zarqa, Jordan. E-mail: malauthman@zu.edu.jo*
- *Mohammad Al Khaldy, Al Khawarizmi University Technical College, Amman, Jordan, E-mail: m.khaldy@khwawarizmi.edu.jo*

4416on the network's life. The study in [3] focuses on energy consumption based on a typical sensor node's hardware parts. Authors classify a radio subsystem, a sensor subsystem, a power supply, and a processing subsystem as four primary elements of the sensor nodes. Authors in [4][5] have taken into consideration in recent research the problems with energy-aware broadcast / multicast in order to classify algorithms into two classes: the maximum lifetime broadcast/multicast (MLB / MLM) problem in wireless ad-hoc networks ; and the (MLB / MEM) problem in terms of minimal energy broadcast / multicast. The primary energy-aware metrics reduce the overall energy exhaustion of all the node engaged in the multi-cast session and boost the processing time until the battery depletion of the first multi-cast node occurs. The authors of [6] clarify the difficulties involved in designing energy-efficient WSN protocols for the Medium Access Control (MAC). Furthermore, it discusses a few WSN MAC protocols confirming their pros and cons. Three types of issues are presented in [7] by the authors: inner platform and operating system, communications stack protocol, network services, delivery, and implementation. A protocol for energy balance has been proposed by the authors in [8] and is accomplished by clustering between sensor nodes. The clustering strategy involves the remaining power of the nodes and sensor node distances into account. Authors in [9], the research addressed the distinctions between many well-known protocols like Low energy adaptive clustering hierarchy [LEACH], Threshold sensitive energy-efficient sensor network protocol [TEEN], Adaptive Periodic Threshold Sensitive Energy Efficient Sensor Network Protocol (APTEEN), Hybrid, Energy-Efficient Distributed Clustering (HEED), Power-Efficient Gathering in Sensor Information Systems (PEGASIS). In order to search for solid techniques to solve the issue of deployment from a realistic view, the authors suggested the Relay Node Placement Problem (RNPP) protocol [10] in WSNs of two multi-objective formulations, which includes energy cost and average sensitivity and network reliability.

### 2.2 Routing Algorithms
The literature on WSN's routing protocols contains several studies presented below. Authors in [11] Presented a study of wireless sensor network routing protocols, this survey classifies network-based routing techniques into three classes: the hierarchical, flat and location-based routing protocols. These protocols are also classified as negotiation-based, multi-path, query-based and QoS-based protocol routing methods. Authors in [12] have presented a classification of protocol routing and design problems, as well as a protocol for routing based on their features and mechanisms, without giving any information about each of the protocols described. Authors in [13] have suggested an Environment-fusion multipath routing protocol (EFMRP) multi-path environment-fusion routing, and this includes a potential field modelling method for instructing data packets to reach the sink. Maintenance, traffic distribution and retreat mechanism design. In [14], researchers discussed the issue of the energy balance and minimisation in WSN's energy consumption. Authors have developed an energy distributed area algorithm to select high energy nodes dynamically as source routing nodes which serve to calculate the route for other common nodes. In addition, the authors present the effective distance as a criterion for the optimal transmission distance and suggest an efficient ant colony optimisation algorithm for finding an ideal origin route for each specific node.

### 2.3 Design and Architecture
Authors in [15] discussed wireless sensor network design issues and methods by illustrating the physical restrictions and proposed protocols for all network stack layers of sensor nodes, and they also talked about the feasible implementations of WSNs. A few sensor networking protocols mentioned in the [16] study and classified them into data-centric, hierarchical and location-based protocols), but the article does not concentrate on energy-efficient strategies; however, the paper talks about WSN routing

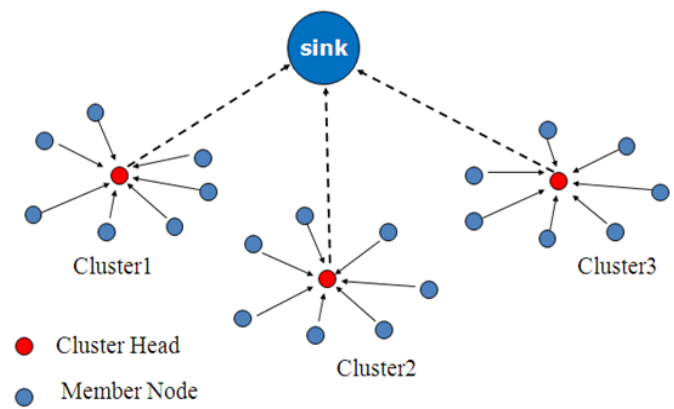

protocols.

WSN Hierarchical routing protocols to accomplish energy consumption effectiveness have already been provided by cluster architectures In general, the sensor nodes are divided into groups; each group called cluster as shown in Figure 1, A chosen node or leader known as the Cluster Head (CH) is maintained for each cluster.

**Figure 1. Clustering in the wireless sensor network.**

Clustering was used by WSNs to increase the scalability of the network, the sharing of resources and efficient use of resources, giving stability to the network topology and saving energy. Using clustering method, energy drain reduces by limiting the communication variety since all nodes are communicating with a CH acting as a local sink requiring lower energy, Because there is a narrower range between the sensors and the local sink than the range between the sensors and the worldwide sink, the CH transfers sensed information to the worldwide sink to decrease energy usage, and allows some nodes to power-off inside the cluster.

**Objectives Of Clustering**
a. More Scalability: In the design of routing protocols for the WSNs, scalability is very important and critical. We might say that the routing protocol is effective if it is flexible and changes in network topology from time to time with good performance, WSNs can be scaled if after design we can add network nodes [17].

b. Data aggregation: The data aggregation collects helpful information from nodes, which is one of the key characteristics for energy savings in WSN's, to prolong network life, so optimisation is required. To achieve effective data aggregation, energy saving must be achieved, and the unnecessary repeated information removed [ 12 ].

The overall data aggregation algorithm is shown in Figure 2.

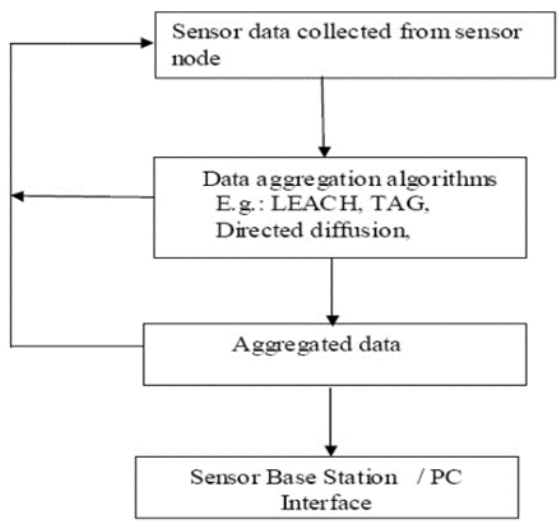

**Figure 2.** The general architecture of data aggregation algorithm.

c. Collision Avoidance: Can influence the entire system efficiency, so that we should prevent collision to improve the network lifetime Because excessive collisions lead to loss, resource waste, and retransmission, The result has been costs and latency, and intensive WSNs have worsened [18].

d. Latency Reduction: The WSNs are divided into clusters, each cluster has own cluster head, and only the CHs transmit the information to the sink, Since the information transfer method is removed from the cluster, therefore the collision and latency are reduced. In addition, hop-by-hop information transmission is often performed using flooding as a flat routing system, while the clustering routing system is used by only CHs, that can decrease hops to the base station from the information source; therefore, latency is reduced [19].

e. Load Balancing: Performing clusters equal in size is important for improved network life as precocious energy consumption of cluster heads is avoided and we should balance the load between CHs in order to achieve the required performance, where they are selected from available sensors [19][20] [21]. When clusters have the same number of nodes, data from individual CHs are ready for further work in the base station at the same time.

f. Fault-Tolerance: The nodes of WSN can have to operate under a tough circumstance, so they generally experience drainage, transfer errors or delay, crash of hardware, malicious attacks, etc. Therefore, fault tolerance is required to prevent the loss of significant packets [21]. Re-clustering is the intuitive way to get back from the failure of the cluster head. [19] [20] [21].

g. Guarantee of Connectivity: WSN CHs transmit the information to the base station with a single hop or multi-hop routing, In order to guarantee that information are passed on to the BS effectively, we have to determine if each node are linked to the next-hop node, If the node is not able to communicate information to BS because no other node can communicate it becomes isolated. [20].

### 4)　Artificial Intelligence

Authors in [22] present an artificial intelligence algorithm which routes the packet effectively across the network without the heavy consumption of batteries, Two protocols based on KNN and ANN artificial intelligence algorithms were presented in the paper; the authors demonstrate a better system performance by using the proposed algorithm. Authors in [23] introduced the Best First Search artificial intelligence Algorithm (BFS) the In selecting a next-hop communication path, protocol has used a multivariable heuristic function, the protocol aimed to Maximize the lifetime of the wireless sensor network. In addition, system reliability and the average delivery rate of packets to be improved, simulation compared with OLSR and LASeR protocols. A protocol was proposed by authors in [24] on the basis of genetic algorithms for energy-aware multipath routing techniques, authors have introduced a cost function which takes into account the distance between the centre and the event area and base station together with the node residual energy for choosing CH. Authors were presented in the proposed algorithm [25] to determine the optimum number and position of the deployment of the relay Node based on the algorithm for particle optimisation. The results of the algorithm proposed have been revealed to reduce the number of feasible primary conditions and thus the cost of network creation.

### 5)　Sink mobility

A new sink mobility algorithm has been provided in [26] article which includes network coverage for routing data packets to mobile sinks in a partition of various sink wireless sensor networks. The authors showed that network lifetime could be improved through the deployment of a sink moving in line with the designs specified in the algorithm. Due to the shortening of source range, energy consumption while communicating with information is decreased. Location-Aware Routing for Controlled Mobile Sinks (LARCMS) [27] is proposed by authors to assist minimising reporting delay, improve network life, handle sink position updates and provide a consistent power consumption. The technology suggested utilises two portable sinks for the collection of information for a predefined trajectory and offers better outcomes than current technology.

## 3. CONCLUSIONS

Various apps used WSNs, but there is an issue of power consumption, so we introduced various wireless network protocols for how to increase the lifetime of WSNs, we have summarised and categorised various methods for addressing and classifying energy efficiency challenges in WSNs. We have indicated for each class of methods which energy waste source it reduces.

## AUTHORS PROFILE

**Ghassan Samara:** Holds BSc. and MSc. in Computer Science, and PhD in Computer Networks. He obtained his Ph.D, from Universiti Sains Malaysia (USM) in 2012. His field of specialization is Cryptography, Authentication, Computer Networks, Computer Data and Network Security,

Wireless Networks, Vehicular Networks, Inter-vehicle Networks, Car to Car Communication, Certificates, Certificate Revocation, QoS, Emergency Safety Systems. Currently, Dr. Samara is the assistant professor at Zarqa University, Jordan.

**Ghadeer Albesani:** Is currently working at The International Arab Journal of Information Technology (IAJIT). She received her BSc in Software engineering from the Zarqa University in 2014. She received MSc of computer science from Zarqa University in 2019.

**Mohammed Alauthman:** Received his PhD degree from Northumbria University Newcastle, UK in 2016. He received a B.Sc. degree in Computer Science from Hashemite University, Jordan, in 2002, and received M.Sc. degrees in Computer Science from Amman Arab University, Jordan, in 2004. Currently, he is Assistant Professor and senior lecturer at department of computer science, Zarqa University, Jordan. His research interests include cyber-security, Cyber Forensics, advanced machine learning and data science applications.

**Mohammad Al Khaldy:** Assistant Professor with a demonstrated history of working in the higher education.
Skilled in lecturing, machine learning, statistical modeling, R, Python, data science and analytical skills.
Strong research professional towards a doctor of philosophy (PhD) focused in predictive modeling for heart failures at the University of Hull.